\documentclass[fleqn,usenatbib]{mnras}

\usepackage{newtxtext,newtxmath}

\usepackage[T1]{fontenc}
\usepackage{xcolor}
\DeclareRobustCommand{\VAN}[3]{#2}
\let\VANthebibliography\thebibliography
\def\thebibliography{\DeclareRobustCommand{\VAN}[3]{##3}\VANthebibliography}

\newcommand{\orcid}[1]{\href{https://orcid.org/#1}{\textcolor[HTML]{A6CE39}{\includegraphics[height=1.7ex]{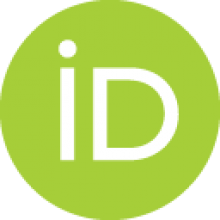}}}}

\allowdisplaybreaks

\usepackage{graphicx}	
\usepackage{amsmath}	
\usepackage{widetext}

\newcommand{\vela}{\texttt{Vela.jl}}
\newcommand{\pint}{\texttt{PINT}}

\newcommand{\enterprise}{\texttt{ENTERPRISE}}
\newcommand{\tempo}{\texttt{tempo}}
\newcommand{\tempotwo}{\texttt{tempo2}}
\newcommand{\temponest}{\texttt{TEMPONEST}}

\newcommand{\updated}[1]{
#1
}

\title[DMGP for wideband timing]{Gaussian process representation of dispersion measure noise in pulsar wideband datasets}

\author[Abhimanyu Susobhanan]{
Abhimanyu Susobhanan\textsuperscript{\orcid{0000-0002-2820-0931}}$^{1}$\thanks{E-mail: abhimanyu.susobhanan@aei.mpg.de},
Rutger van~Haasteren\textsuperscript{\orcid{0000-0002-6428-2620}}$^{1}$
\\
$^{1}$Max-Planck-Institut f{\"u}r Gravitationsphysik (Albert-Einstein-Institut), Leibniz Universit{\"a}t Hannover, Callinstra{\ss}e 38, 30167 Hannover, Deutschland
}

\date{Accepted XXX. Received YYY; in original form ZZZ}

\pubyear{\the\year{}}

\begin{document}
\label{firstpage}
\pagerange{\pageref{firstpage}--\pageref{lastpage}}
\maketitle

\begin{abstract}
The ionized interstellar medium disperses pulsar radio signals, resulting in a stochastic time-variable delay known as the dispersion measure (DM) noise.
In the wideband paradigm of pulsar timing, we measure a DM together with a time of arrival from a pulsar observation to handle frequency-dependent profile evolution, interstellar scintillation, and radio frequency interference more robustly, and to reduce data volumes.
In this paper, we derive a method to incorporate arbitrary models of DM variation, including Gaussian process models, in pulsar timing and noise analysis and pulsar timing array analysis.
This generalizes the existing method for handling DM noise in wideband datasets. 
\end{abstract}

\begin{keywords}
pulsars: general -- methods: data analysis
\end{keywords}



\section{Introduction} 
\label{sec:intro}

Pulsars, rotating neutron stars emitting electromagnetic radiation that appears as periodic pulses to terrestrial observers, are some of the most rotationally stable objects in the universe \citep{LorimerKramer2012}.
Radio waves emitted by pulsars are dispersed as they travel through the ionized interstellar medium (ISM), inducing delays that are proportional to the electron column densities along their lines of sight (also known as the dispersion measure, DM) and inversely proportional to the square of the observing frequency \citep{BackerHellings1986}.
The DM varies stochastically as a function of time due to the dynamic nature of the ISM and the fact that the pulsar and the Earth are moving relative to each other \citep[e.g.][]{DonnerVerbiest+2020}.
Other astrophysical effects that influence the times of arrival (TOAs) of pulsar pulses include the motion of the Earth through the solar system \citep{EdwardsHobbsManchester2006}, the motion of the pulsar in a binary system \citep{DamourDeruelle1986}, the proper motion of the pulsar, and gravitational waves passing across the line of sight \citep{EstabrookWahlquist1975}.
Pulsar timing is the technique of tracking a pulsar's rotation by measuring the TOAs of its pulses, allowing the pulsar to be used as a celestial clock \citep{HobbsEdwardsManchester2006}.
Pulsar timing has been used to probe a wide range of phenomena, from neutron star equation of state \citep[e.g.][]{CromartieFonseca+2020} to nanohertz gravitational waves \citep[e.g.][]{AgazieAntoniadis+2024}.

Radio pulsar timing is conventionally performed by folding the frequency-resolved pulsar light curve using the known rotational period to obtain a frequency-resolved integrated pulse profile.
The integrated pulse profile is then split into multiple frequency sub-bands, and a TOA is measured independently from each sub-band by cross-correlating the profile against a noise-free template \citep{Taylor1992}.
While this paradigm, known as narrowband timing, is widely used, it has two major drawbacks: it produces a large number of TOAs leading to large data volumes, and it does not adequately account for frequency-dependent profile evolution \citep{HankinsRickett1986} in many cases\footnote{{In principle, frequency-dependent profile evolution can be characterized accurately in narrowband timing by using a sufficiently large number of frequency sub-bands. However, this is not always possible in practice since the per-sub-band signal-to-noise ratio decreases as the number of sub-bands increases. Further, using a large number of sub-bands leads to larger data volumes.
}}.
The more recently developed wideband timing technique handles these issues by treating the frequency-resolved integrated pulse profile as a single two-dimensional entity known as a portrait and simultaneously measures a single TOA and a single dispersion measure (DM) by cross-correlating it against a two-dimensional template \citep{LiuDesvignes+2014,PennucciDemorestRansom2014}.
The two-dimensional template is usually derived from data using a principal component analysis-based algorithm \citep{Pennucci2019}.
An extension of the wideband method for combining simultaneous multi-frequency observations during TOA and DM estimation was presented in \citet{PaladiDwivedi+2023}.
Remarkable applications of this method include \citet{AlamArzoumanian+2021}, \citet{TarafdarNobleson+2022},  \citet{CuryloPennucci+2023}, \citet{AgazieAlam+2023}, and \citet{TanFonseca+2024}.

Since wideband timing is a relatively new method, data analysis techniques and software tools for handling wideband datasets are still being developed.
Wideband TOA and DM measurements are performed using the \texttt{PulsePortraiture} package \citep{PennucciDemorestRansom2014,Pennucci2019}.
Interactive timing of wideband datasets can be performed using the \tempo{} \citep{NiceDemorest+2015} and \pint{} \citep{LuoRansom+2021,SusobhananKaplan+2024} packages.
Bayesian noise analysis of wideband datasets assuming a linearized timing model (commonly known as single-pulsar noise analysis, SPNA) is available in a limited capacity via the \enterprise{} package \citep{EllisVallisneri+2020,JohnsonMeyers+2024}, 
and Bayesian timing \& noise analysis using the full non-linear timing model (commonly known as single-pulsar noise \& timing analysis, SPNTA) can be performed using the \vela{} package \citep{Susobhanan2025a,Susobhanan2025b}\footnote{{\vela{} is new SPNTA package written in Julia and Python supporting both narrowband and wideband paradigms. It is designed to be efficient, flexible, and easy-to-use, and is available at \url{https://abhisrkckl.github.io/Vela.jl/}.}}.
{The \tempotwo{} package \citep{HobbsEdwardsManchester2006} used for interactive pulsar timing and the \temponest{} package \citep{LentatiAlexander+2014} used for SPNTA of narrowband datasets do not currently support the wideband timing paradigm.}
\enterprise{} also enables pulsar timing array \citep[PTA:][]{FosterBacker1990} data analysis using wideband datasets, which combines multiple pulsar datasets to search for signals that are correlated across pulsars. 

The wideband timing and noise analyses available in the literature have generally employed a piecewise-constant model for DM variations, known as the DMX model \citep{ArzoumanianBrazier+2015,AlamArzoumanian+2021}, and this is the only DM variability model currently available in \enterprise{} for wideband datasets.
However, recent studies have shown that Gaussian process (GP) models \citep{van_HaasterenVallisneri2014b} for interstellar DM (commonly known as DMGP) and solar wind variability may be more suitable in many cases \citep{LarsenMingarelli+2024,IraciChalumeau+2024,SusarlaChalumeau+2024}.
{In particular, DMGP models are better for modeling long-timescale DM variations since they manifestly incorporate the time correlations, whereas single-epoch DM events like the solar coronal mass ejection reported in \citet{KrishnakumarManoharan+2021} are better modeled using DMX; combining both models may be advantageous in some cases.
Further, irregular infinitely wide prior distributions are usually used for DMX parameters \citep{AlamArzoumanian+2021}, whereas DMGP amplitudes are usually constrained by a spectral model \citep{LentatiAlexander+2014}. 
This makes the DMX model unsuitable in the absence of low-frequency or high-bandwidth observations where the DMs are poorly constrained by the data.}

While \vela{} provides the DMGP model in the context of SPNTA, it is also advantageous to incorporate it in SPNA since the latter is usually computationally less expensive.
SPNA can also be generalized to include correlations between different pulsars, such as those produced by the stochastic gravitational wave background \citep[GWB:][]{HellingsDowns1983}, in a computationally tractable way \citep{van_HaasterenLevin+2009}.
In this paper, we develop a formalism to incorporate the DMGP model in SPNA and PTA analysis from first principles.

This paper is arranged as follows. 
We provide a brief overview of the wideband timing \& noise model, used for SPNTA, in Section \ref{sec:timing-model}.
We describe the linearization of the wideband timing model and the analytic marginalization of its posterior distribution, used for SPNA, in Section \ref{sec:timing-model}.
In Section \ref{sec:example}, we demonstrate our results using a simulated dataset.
Finally, we summarize our results in Section \ref{sec:summary}.

\section{The wideband timing \& noise model} 
\label{sec:timing-model}

In the wideband timing paradigm, each observation yields a TOA-DM pair $\left(t_{i},d_{i}\right)$ with uncertainties $\left(\sigma_{i},\epsilon_{i}\right)$ measured simultaneously at a fiducial observing frequency $\nu_i$ within the observing band \citep{PennucciDemorestRansom2014,LiuDesvignes+2014}. 
The measurement covariance $\left\langle t_{i}d_{i}\right\rangle$ is usually made to vanish by adjusting the fiducial observing frequency $\nu_i$ \citep{PennucciDemorestRansom2014}.
The time of emission $\tau_{i}$ can be estimated from the TOA up to an additive constant using the expression \citep{HobbsEdwardsManchester2006,EdwardsHobbsManchester2006}
\begin{align}
    \tau_{i}&=t_{i}-\Delta_{\text{clock}}(t_{i})-\Delta_{\text{jump};i}-\Delta_{\odot}(t_{i})-\Delta_{\text{DM}}(t_{i},\nu_{i})-\Delta_{\text{GW}}(t_{i})\nonumber\\
    &\quad-\Delta_{\text{B}}(t_{i})
    -\mathcal{N}_{\text{R};i}\,,
    \label{eq:delays}
\end{align}
where 
$\Delta_{\text{clock}}$ is a clock correction that converts $t_{i}$ from the observatory timescale to a timescale defined at the solar system barycenter, 
$\Delta_{\text{jump}}$ contains instrumental delays introduced by the observing system and the data reduction pipeline,
$\Delta_{\odot}$ represents the delays caused by the solar system motion,
$\Delta_{\text{DM}}$ represents the dispersion delay caused by the ionized interstellar medium and the solar wind,
$\Delta_{\text{GW}}$ represents a delay induced by passing gravitational waves \citep{EstabrookWahlquist1975}, 
$\Delta_{\text{B}}$ represents the delays caused by the binary motion of the pulsar \citep{DamourDeruelle1986}, and
$\mathcal{N}_{\text{R};i}$ is an uncorrelated (white) Gaussian noise process representing the radiometer noise \citep{LorimerKramer2012}. 
The rotational phase $\phi_{i}$ corresponding to the $i$th TOA is given by \citep{HobbsEdwardsManchester2006}
\begin{equation}
\phi_{i}=\phi_{0}+\sum_{j=1}^{N_{F}}\frac{F_{j}(\tau_{i}-\tau_{0})^{j+1}}{(j+1)!}+\phi_{\text{glitch}}(\tau_{i})+\phi_{\text{SN}}(\tau_{i})+\mathcal{N}_{\text{jitter};i}\,,
\label{eq:phase}
\end{equation}
where $\phi_{0}$ is an initial phase, $F_{j}$ represent the rotational frequency and its derivatives, $\phi_{\text{glitch}}$ represents a phase correction due to glitches, $\phi_{\text{SN}}$ represents the slow stochastic wandering of the rotational phase known as the spin noise \citep{ShannonCordes2010}, and $\mathcal{N}_{\text{jitter};i}$ is a time-uncorrelated stochastic process representing pulse jitter \citep{ParthasarathyBailes+2021}.
{Note that the above expressions only incorporate the most commonly encountered effects in pulsar timing; other effects may include, but are not limited to, three-body interactions \citep[e.g.,][]{RansomStairs+2014}, acceleration due to globular cluster potential \citep[e.g.,][]{DaiJohnston+2023}, profile shape change events \citep[e.g.,][]{JenningsCordes+2024}, etc. 
Further,} we have \textit{not} considered interstellar scattering \citep{HembergerStinebring2008} in the above expressions since the correct way to incorporate it in wideband timing is not yet understood.

The timing residual $r_{i}$ is given by \citep{HobbsEdwardsManchester2006}
\begin{equation}
r_{i}=\frac{\phi_{i}-N[\phi_{i}]}{\bar{F}_{i}}\,, 
\label{eq:toa-res}
\end{equation}
where $N[\phi_{i}]$ represents the integer closest to $\phi_{i}$,
and $\bar{F}_{i}=d\phi_{i}/dt_{i}$ is the topocentric frequency of the pulsar. 
Similarly, the DM residual $\delta_{i}$ is given by \citep{PennucciDemorestRansom2014,AlamArzoumanian+2021}
\begin{align}
\delta_{i}&=d_{i}-\sum_{j=1}^{N_{D}}\frac{D_{j}(t_{i}-t_{0})^{j}}{j!}-\mathcal{D}_{\text{DMN}}(t_{i})-\mathcal{D}_{\text{SW}}(t_{i})-\mathcal{D}'_{\text{jump};i}-\mathcal{M}_{i}\nonumber\\
&=d_{i}-\mathcal{D}(t_{i})-\mathcal{D}'_{\text{jump};i}-\mathcal{M}_{i}\,,
\label{eq:dm-res}
\end{align}
where $D_{j}$ represents the interstellar DM and its derivatives,
$\mathcal{D}_{\text{DMN}}$ represents the stochastic variations in the interstellar DM known as the DM noise, 
$\mathcal{D}_{\text{SW}}$ represents the solar wind DM, $\mathcal{D}'_{\text{jump}}$ represent system-dependent wideband DM offsets known as DMJUMPs, 
and $\mathcal{M}_{i}$ represents a white noise process with contributions from radiometer noise and pulse jitter. 
$\mathcal{D}$ represents the astrophysical DM, which is the sum of all DM contributions except $\mathcal{D}'_{\text{jump}}$. 
$\mathcal{D}$ is related to $\Delta_{\text{DM}}$ as 
\begin{equation}
    \Delta_{\text{DM}}(t_{i})=\mathcal{K}\mathcal{D}(t_{i})\nu_{i}^{-2}\,,
    \label{eq:dm-delay}
\end{equation}
where $\mathcal{K}\approx 4.15 \times 10^3$ MHz$^2$ s pc$^{-1}$ cm$^3$  is the dispersion constant.

~

Considering a wideband dataset containing $N_{\text{toa}}$ measurements, we define a $2N_{\text{toa}}$-dimensional residual vector $\textbf{y}$ as 
\begin{equation}
    \textbf{y}^T = [r_1, ..., r_{N_{\text{toa}}}, \delta_1, ..., \delta_{N_{\text{toa}}}]\,,
    \label{eq:res-vec}
\end{equation}
and a $2N_{\text{toa}}\times 2N_{\text{toa}}$ covariance matrix $\textbf{N}$ as 
\begin{equation}
    \textbf{N} = \text{diag}[\varsigma_1^2, ..., \varsigma^2_{N_{\text{toa}}}, \varepsilon^2_1, ..., \varepsilon^2_{N_{\text{toa}}}]\,,
    \label{eq:white-cov}
\end{equation}
where $\varsigma_i$ and $\varepsilon_i$ are the scaled TOA and DM uncertainties given by
\begin{subequations}
\begin{align}
\varsigma_{i}^{2}&=E_i^{2}\left(\sigma_{i}^{2}+Q_i^{2}\right)\,,\\
\varepsilon_{i}^{2}&=\mathcal{E}_{i}^{2}\left(\epsilon_{i}^{2}+\mathcal{Q}_{i}^{2}\right)\,.
\end{align}
\end{subequations}
The quantities $E$, $Q$, $\mathcal{E}$, and $\mathcal{Q}$ modifying the uncertainties are known as EFACs, EQUADs, DMEFACs, and DMEQUADs, and are observing system-dependent \citep{LentatiAlexander+2014,AlamArzoumanian+2021}.
If the measurement covariances $\left\langle t_{i}d_{i}\right\rangle$ are non-zero, they can be included in $\textbf{N}$ as non-diagonal elements.

These definitions allow us to write the wideband timing likelihood function as
\begin{equation}
    L = \frac{1}{\sqrt{\det [2\pi\textbf{N}]}} \exp\left[-\frac{1}{2} \textbf{y}^T \textbf{N}^{-1} \textbf{y}\right] \,.
    \label{eq:lnlike-full}
\end{equation}

\section{The linearized timing model} 
\label{sec:linear-timing-model}

In a frequentist setting, an optimal timing solution is found by maximizing $\ln L$ over the model parameters appearing in equations \eqref{eq:delays}-\eqref{eq:lnlike-full} \citep{LuoRansom+2021,SusobhananKaplan+2024}.
Although the residuals $\textbf{y}$ are generally non-linear functions of the timing model parameters, they can be approximated as linear functions in the vicinity of the maximum likelihood point in the parameter space \citep[][]{DamourDeruelle1986,HobbsEdwardsManchester2006,vanhaasterenPlacingLimitsStochastic2011}. 
i.e., 
\begin{equation}
\textbf{y} = \bar{\textbf{y}} + \textbf{M} \textbf{a}\,,
\label{eq:linear-tm}
\end{equation}
where $\bar{\textbf{y}}$ represents the residuals at a reference point, 
$\textbf{a}$ is a $p$-dimensional vector containing deviations of the timing model parameters from their values at the reference point, and
$\textbf{M}$ is a $2N_{\text{toa}}\times p$ design matrix containing partial derivatives of the residuals $y_\alpha$ with respect to the parameter deviations $a_\beta$ (i.e., $M_{\alpha\beta}=\frac{\partial y_\alpha}{\partial a_\beta}$).
An example structure for $\textbf{M}$ is shown below.
\begin{equation}
    \text{\textbf{M}}^{T}=\begin{bmatrix}\frac{\partial r_{1}}{\partial\phi_{0}} & ... & \frac{\partial r_{N_{\text{toa}}}}{\partial\phi_{0}} & 0 & ... & 0\\
\frac{\partial r_{1}}{\partial F_{0}} & ... & \frac{\partial r_{N_{\text{toa}}}}{\partial F_{0}} & 0 & ... & 0\\
\frac{\partial r_{1}}{\partial F_{1}} & ... & \frac{\partial r_{N_{\text{toa}}}}{\partial F_{1}} & 0 & ... & 0\\
\vdots & \vdots & \vdots & \vdots & \vdots & \vdots\\
\frac{\partial r_{1}}{\partial a_{1}^{\text{SN}}} & ... & \frac{\partial r_{N_{\text{toa}}}}{\partial a_{1}^{\text{SN}}} & 0 & ... & 0\\
\frac{\partial r_{1}}{\partial b_{1}^{\text{SN}}} & ... & \frac{\partial r_{N_{\text{toa}}}}{\partial b_{1}^{\text{SN}}} & 0 & ... & 0\\
\vdots & \vdots & \vdots & \vdots & \vdots & \vdots\\
\frac{\partial r_{1}}{\partial D_{0}} & ... & \frac{\partial r_{N_{\text{toa}}}}{\partial D_{0}} & \frac{\partial\delta_{1}}{\partial D_{0}} & ... & \frac{\partial\delta_{N_{\text{toa}}}}{\partial D_{0}}\\
\frac{\partial r_{1}}{\partial D_{1}} & ... & \frac{\partial r_{N_{\text{toa}}}}{\partial D_{1}} & \frac{\partial\delta_{1}}{\partial D_{1}} & ... & \frac{\partial\delta_{N_{\text{toa}}}}{\partial D_{1}}\\
\vdots & \vdots & \vdots & \vdots & \vdots & \vdots\\
\frac{\partial r_{1}}{\partial a_{1}^{\text{DMN}}} & ... & \frac{\partial r_{N_{\text{toa}}}}{\partial a_{1}^{\text{DMN}}} & \frac{\partial\delta_{1}}{\partial a_{1}^{\text{DMN}}} & ... & \frac{\partial\delta_{N_{\text{toa}}}}{\partial a_{1}^{\text{DMN}}}\\
\frac{\partial r_{1}}{\partial b_{1}^{\text{DMN}}} & ... & \frac{\partial r_{N_{\text{toa}}}}{\partial b_{1}^{\text{DMN}}} & \frac{\partial\delta_{1}}{\partial b_{1}^{\text{DMN}}} & ... & \frac{\partial\delta_{N_{\text{toa}}}}{\partial b_{1}^{\text{DMN}}}\\
\vdots & \vdots & \vdots & \vdots & \vdots & \vdots\\
0 & ... & 0 & \frac{\partial\delta_{1}}{\partial J_{1}} & ... & \frac{\partial\delta_{N_{\text{toa}}}}{\partial J_{1}}\\
\vdots & \vdots & \vdots & \vdots & \vdots & \vdots
\end{bmatrix}\,.
\label{eq:design-matrix}
\end{equation}
Here, $J_i$ are system-dependent DM jumps appearing in $\mathcal{D}'_{\text{jump}}$, and $D_0$ and $D_1$ are Taylor series coefficients that determine the long-term evolution of the DM.
We have assumed that the spin noise and DM noise are well-approximated by the Fourier series expressions
\begin{subequations}
\begin{align}
    \phi_{\text{SN}}(t_{i})&=\sum_{j=1}^{N_{\text{harm}}}\left\{ a_{i}^{\text{SN}}\cos\left(2\pi jT_{\text{span}}^{-1}\left(t_{i}-t_{0}\right)\right)\right.\nonumber\\
    &\qquad\left.+\,b_{i}^{\text{SN}}\sin\left(2\pi jT_{\text{span}}^{-1}\left(t_{i}-t_{0}\right)\right)\right\} \,, \label{eq:sn-fourier} \\
    \mathcal{D}_{\text{DMN}}(t_{i})&=\sum_{j=1}^{N_{\text{harm}}}\left\{ a_{i}^{\text{DMN}}\cos\left(2\pi jT_{\text{span}}^{-1}\left(t_{i}-t_{0}\right)\right)\right.\nonumber\\&\qquad\left.+\,b_{i}^{\text{DMN}}\sin\left(2\pi jT_{\text{span}}^{-1}\left(t_{i}-t_{0}\right)\right)\right\} \,,
\label{eq:dmn-fourier}
\end{align}
\end{subequations}
with Fourier amplitudes $a_{i}^{\text{SN}}$, $b_{i}^{\text{SN}}$, $a_{i}^{\text{DMN}}$, and $b_{i}^{\text{DMN}}$ \citep{LentatiAlexander+2014}, where
$T_{\text{span}}$ is the total time span of the dataset.

~

Assuming that the best-fit residual vector $\bar{\textbf{y}}$ only contains Gaussian white noise, the likelihood function can be written in this linear regime as
\begin{equation}
    L = \frac{1}{\sqrt{\det[2\pi\textbf{N}]}}\exp\left[-\frac{1}{2}(\textbf{y}-\textbf{M}\textbf{a})^{T}\textbf{N}^{-1}(\textbf{y}-\textbf{M}\textbf{a})\right]\,.
    \label{eq:lnlike-linear}
\end{equation}
It is straightforward to see that the conjugate prior for the parameter deviations $\textbf{a}$ is a multivariate Gaussian distribution.
In particular, following \citet{van_HaasterenLevin+2009}, \citet{van_HaasterenLevin2013}, and \citet{LentatiAlexander+2013}, we choose a Gaussian prior distribution with zero mean and a covariance matrix $\boldsymbol{\Phi}$ defined as a function of parameters \textbf{A}:
\begin{equation}
     \Pi[\textbf{a}|\textbf{A}] = \frac{1}{\sqrt{\det[2\pi\boldsymbol{\Phi}]}}\exp\left[-\frac{1}{2}\textbf{a}^{T}\boldsymbol{\Phi}^{-1}\textbf{a}\right]\,.
    \label{eq:hyperprior}
\end{equation}
It turns out that, in pulsar timing, many of the parameter deviations $\textbf{a}$ are usually strongly constrained by the likelihood function with a very weak dependence on their prior distributions.
Hence, it is customary to use unbounded improper priors for parameters like $\phi_0$, $F_0$, $F_1$, etc, corresponding to infinite diagonal elements in $\boldsymbol{\Phi}$.
On the other hand, Gaussian priors are generally used for the spin noise and DM noise amplitudes, whose variances relate to the spectrum of the noise process by way of the Wiener-Khinchin theorem \citep{van_HaasterenVallisneri2014b,van_HaasterenVallisneri2014a}.
The spin/DM noise spectrum is often parameterized using a function such as a power law.

The log-posterior distribution of the parameter deviations $\textbf{a}$, the white noise parameters $\textbf{b}$ appearing in $\textbf{N}$, and the prior parameters $\textbf{A}$ appearing in $\boldsymbol{\Phi}$ such as the spin noise and DM noise spectral parameters, given a dataset $\mathfrak{D}$ and a timing \& noise model $\mathfrak{M}$ can be written using Bayes theorem as
\begin{align}
P[\textbf{a},\textbf{b},\textbf{A}|\mathfrak{D},\mathfrak{M}] &= \frac{L[\mathfrak{D}|\textbf{a},\textbf{b},\mathfrak{M}]\times\Pi[\textbf{a},\textbf{A}|\mathfrak{M}]\times\Pi[\textbf{b}|\mathfrak{M}]}{Z[\mathfrak{D}|\mathfrak{M}]}\nonumber\\    &=L[\mathfrak{D}|\textbf{a},\textbf{b},\mathfrak{M}]\times\Pi[\textbf{a}|\textbf{A},\mathfrak{M}]\times\Pi[\textbf{A}|\mathfrak{M}]\nonumber\\    &\quad\times\Pi[\textbf{b}|\mathfrak{M}]\,/\,Z[\mathfrak{D}|\mathfrak{M}]\,,
    \label{eq:posterior}
\end{align}
where 
$L[\mathfrak{D}|\textbf{a},\textbf{b},\mathfrak{M}]$ is given by equation \eqref{eq:lnlike-linear},
$\Pi[\textbf{a}|\textbf{A},\mathfrak{M}]$ is given by equation \eqref{eq:hyperprior}, and $Z[\mathfrak{D}|\mathfrak{M}]$ represents the Bayesian evidence.

Since we have imposed conjugate priors on $\textbf{a}$, it is possible to analytically marginalize equation \eqref{eq:posterior} over those parameters.
Following the steps given in {Appendix A of \citet{van_HaasterenLevin+2009}}, we can derive the analytically marginalized posterior as
\begin{align}
    P[\textbf{b},\textbf{A}|\mathfrak{D},\mathfrak{M}]&=\frac{\Pi[\textbf{b}|\mathfrak{M}]\times\Pi[\textbf{A}|\mathfrak{M}]}{Z[\mathfrak{D}|\mathfrak{M}]}\nonumber\\&\quad\times\int d\textbf{a}\;L[\mathfrak{D}|\textbf{a},\textbf{b},\mathfrak{M}]\times\Pi[\textbf{a}|\textbf{A},\mathfrak{M}]\nonumber\\&=
    \frac{\Lambda[\mathfrak{D}|\textbf{b},\textbf{A},\mathfrak{M}]\times\Pi[\textbf{b}|\mathfrak{M}]\times\Pi[\textbf{A}|\mathfrak{M}]}{ Z[\mathfrak{D}|\mathfrak{M}]}\,,
    \label{eq:marg-posterior}
\end{align}
where the marginalized likelihood $\Lambda$ is given by
\begin{equation}
    \Lambda=\frac{1}{\sqrt{\det[2\pi\textbf{C}]}}\exp\left[-\frac{1}{2}\textbf{y}^{T}\textbf{C}^{-1}\textbf{y}\right]\,,
    \label{eq:lnlike-marg}
\end{equation}
with a new covariance matrix
\begin{equation}
    \textbf{C} = \textbf{N} + \textbf{M} \boldsymbol{\Phi} \textbf{M}^T\,.
    \label{eq:cov-marg}
\end{equation}
{Although the example design matrix shown in equation \eqref{eq:design-matrix} only includes a Fourier series representation of the DM along with a Taylor series to account for the lower frequency variations, the formalism derived above is general enough to include any model of DM variations, including piecewise-constant \citep{ArzoumanianBrazier+2015} and constrained spline \citep{KeithColes+2013} models as well as various solar wind models \citep{YouHobbs+2007, YouColes+2012,MadisonCordes+2019,HazbounSimon+2022,SusarlaChalumeau+2024}.}

{A remark is in order regarding the computational cost of evaluating the marginalized likelihood given by equation \eqref{eq:lnlike-marg}, where the Woodbury identity is used to evaluate $\textbf{y}^T \textbf{C}^{-1} \textbf{y}$ and the matrix determinant identity is used to evaluate $\det \textbf{C}$. 
Both of these computations are dominated by the evaluation of $\textbf{M}^T \textbf{N}^{-1} \textbf{M}$, which has a computational complexity of $\mathcal{O}(N_{\text{data}}\,p^2)$, where $N_{\text{data}}=2N_{\text{toa}}$ for wideband data and $p$ is the number of columns in $\textbf{M}$.
It follows that the likelihood evaluation for a narrowband dataset derived from the same observations, with $N_{\text{subband}}$ sub-bands per epoch, will be $N_{\text{subband}}/2$ times slower than its wideband counterpart, everything else being equal.
PTA data analysis using multiple pulsars is usually performed assuming fixed white noise parameters, and $\textbf{M}^T \textbf{N}^{-1} \textbf{M}$ for the entire PTA is evaluated once and cached in this case.
The computation in this case becomes dominated either by the evaluation of $\textbf{M}^T \textbf{N}^{-1} \textbf{y}$ (complexity of $\mathcal{O}( N_{\text{data}} p)$ per pulsar) or by the inversion of $\left(\boldsymbol{\Phi}^{-1}+\textbf{M}^{T}\textbf{N}^{-1}\textbf{M}\right)$ through Cholesky or singular-value decomposition (worst-case complexity of $\mathcal{O}(P^3)$ where $P$ is the total number of marginalized parameters including single-pulsar and common parameters), depending on the specifics of the data and the model.
} 

Note that the above log-likelihood expression contains cross terms between TOA and DM residuals, unlike the expressions given in Appendix B of \citet{AlamArzoumanian+2021}, where the log-likelihood was fully separated into a TOA part and a DM part.
{This difference arises simply because the expressions in Appendix B of \citet{AlamArzoumanian+2021} are \textit{not} marginalized over the DMX parameters or DMJUMPs, and because the measurement covariances $\left\langle t_{i}d_{i}\right\rangle$ are assumed to be 0.}
This is discussed in detail in Appendix \ref{sec:lnlike-marg-dmx}.

{The \texttt{WidebandTimingModel} class in the \enterprise{} package \citep{EllisVallisneri+2020} implements a version of equation \eqref{eq:lnlike-marg} that only supports the piecewise-constant (DMX) model of DM variations.
In Appendix \ref{sec:lnlike-enterprise}, we derive a factorized \enterprise{}-friendly version of equation  \eqref{eq:lnlike-marg} that is a generalization of the current \enterprise{} implementation.}

{Finally, the fitting methods available in \pint{} for wideband datasets currently do not support DMGP models.
In Appendix \ref{sec:maxlike}, we derive a general maximum-likelihood estimator for wideband timing from our equation \eqref{eq:lnlike-linear} that can be implemented in \pint{}.}

\section{Application to a simulated dataset}
\label{sec:example}

We now demonstrate our method by applying it to a simple simulated dataset.
The simulated dataset corresponds to a fictitious isolated pulsar whose parameters are listed in Table \ref{tab:sim-params}.
We simulate 500 uniformly spaced uncorrelated wideband TOA-DM pairs between MJDs 50000 and 60000 each with a TOA uncertainty of 1 $\mu$s and a DM uncertainty of $10^{-4}$ pc/cm$^3$, taken at the Green Bank Telescope at observing frequencies of 500 MHz, 1000 MHz, and 1500 MHz.
The solar system delays were computed using the DE440 solar system ephemeris \citep{ParkFolkner+2021}.
Further, we inject spin and DM noise with power law spectra whose parameters are given in Table \ref{tab:sim-params}.\footnote{The covariance matrix $\textbf{C}_{\text{sn}}$ of spin noise with a power law spectrum is given by equation (10) of \citet{van_HaasterenVallisneri2014b}. We generate a noise realization as $\texttt{L}_{\text{sn}}\texttt{z}$ where $\texttt{L}_{\text{sn}}$ is the Cholesky factor or $\textbf{C}_{\text{sn}}$ and $\texttt{z}$ is a vector containing $N_\text{toa}$ samples from a unit normal distribution. A DM noise realization can be similarly obtained.}

\begin{figure*}
    \centering
    \includegraphics[width=1\linewidth]{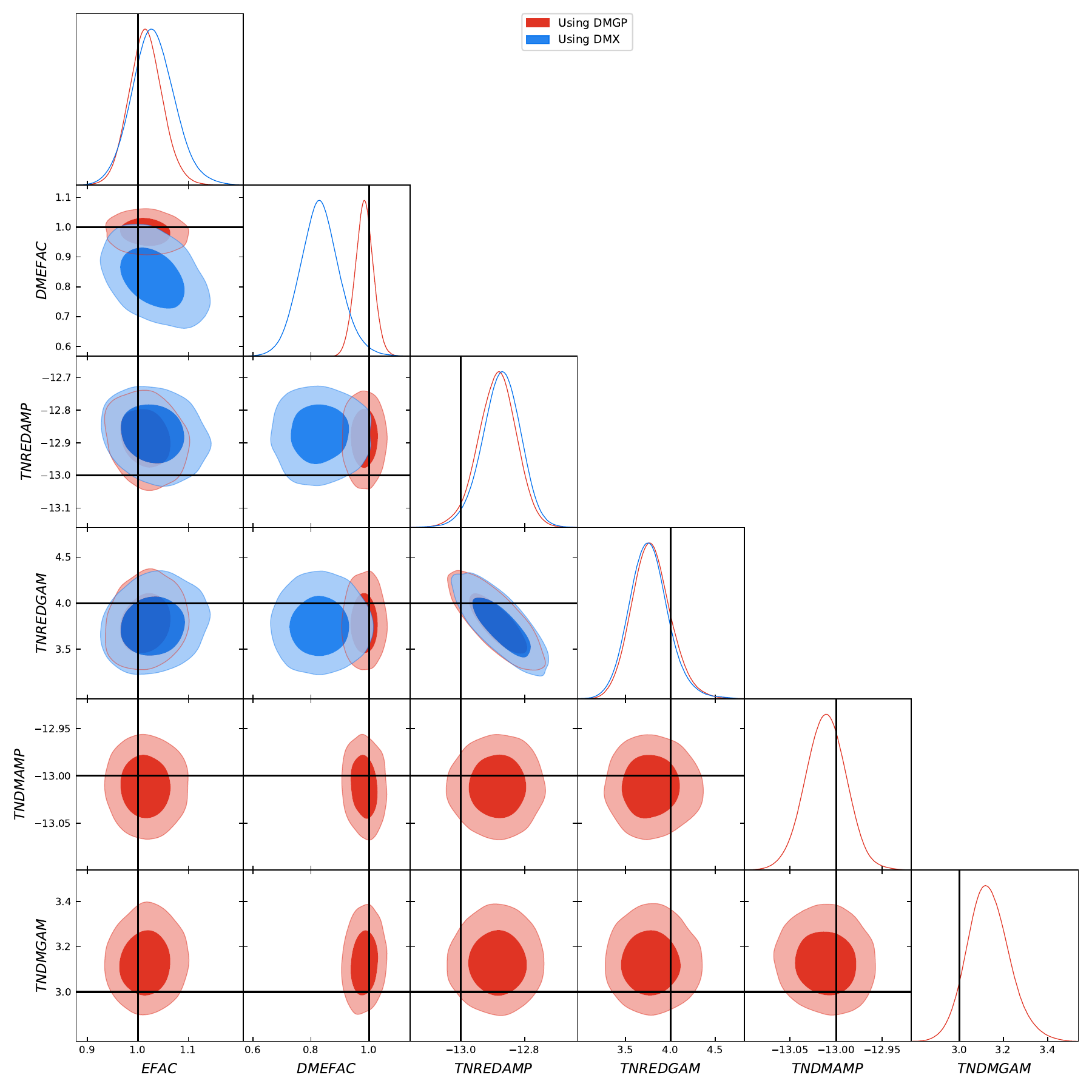}
    \caption{Parameter estimation results for the simulation described in Section \ref{sec:example}.
    \updated{The corner plots show the posterior samples, and the black lines represent the injected parameter values.
    The posterior distribution for the DMGP analysis is shown in red and the posterior distribution for the DMX analysis is shown in blue.
    The parameter estimates are consistent with the injected values within $3\sigma$ uncertainties.
    The pre-fit and post-fit whitened time and DM residuals for the DMGP analysis are shown in Figure \ref{fig:sim-result-res}.}}
    \label{fig:sim-result}
\end{figure*}

\begin{figure*}
    \centering
    \includegraphics[scale=0.7]{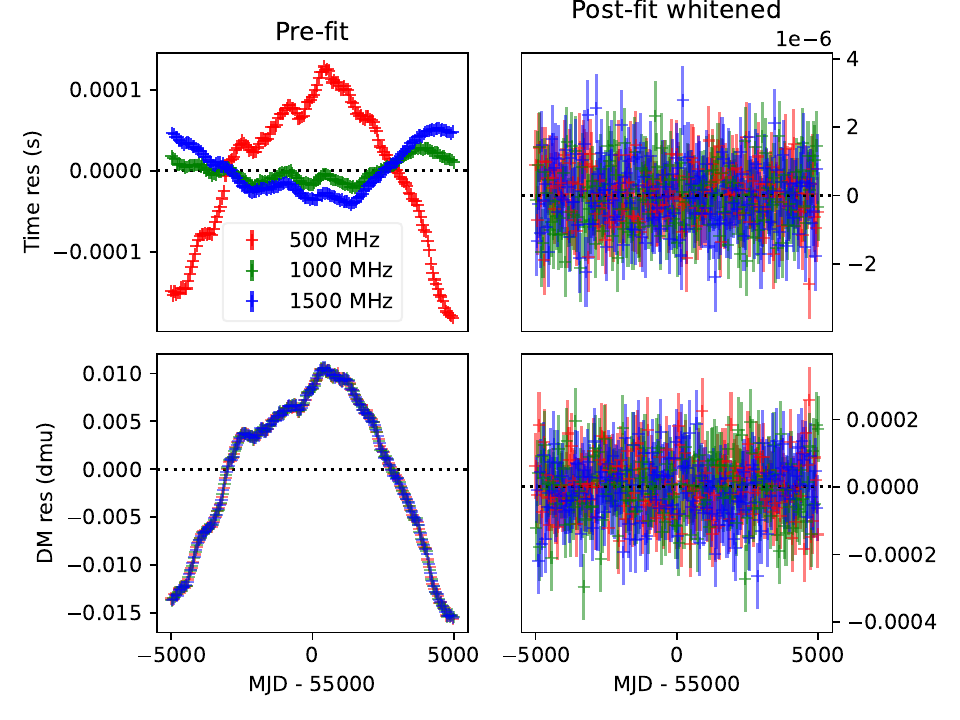}
    \caption{\updated{The pre-fit and whitened time and DM residuals obtained from the simulation study described in Section \ref{sec:example} for the DMGP analysis.
    The colors indicate different observing frequencies.
    The whitened residuals are computed using equations \eqref{eq:ahat}--\eqref{eq:white-res}.
    We see that the residuals have been effectively whitened. 
    The three tracks seen in the top left panel correspond to three observing frequencies.
    They are not simply scaled copies of each other because of the distortion caused by fitting an initial timing model without the correct noise model.}}
    \label{fig:sim-result-res}
\end{figure*}

We implement the likelihood function given in equation \eqref{eq:lnlike-marg} with the help of \pint{}.\footnote{This is not yet available in \enterprise{}.}
We model the spin noise and DM noise as Fourier sums of fundamental frequency $f_1=T_\text{span}^{-1}$, $T_\text{span}=10000$ days, with 120 linearly spaced frequency bins (at $f_1$, $2f_1$, etc). 
Following \citet{van_HaasterenVallisneri2014a}, we also include four logarithmically spaced frequency bins below $f_1$ to better capture the lower frequency components of the spin noise and DM noise (at $f_1/2$, $f_1/4$, $f_1/8$, and $f_1/16$). 
We sample this distribution using the \texttt{emcee} package \citep{Foreman-MackeyHogg2013}, which implements the affine-invariant ensemble sampler algorithm.
The prior distributions used in this analysis are listed in Table \ref{tab:sim-params}.
The posterior distribution and the post-fit residuals obtained from this exercise are plotted in \updated{Figures  \ref{fig:sim-result} and \ref{fig:sim-result-res} respectively}.
We see that all parameter estimates are consistent with injected values within $2\sigma$ uncertainties and that the residuals have been effectively whitened.
We repeated this analysis with a smaller number of linearly spaced frequency bins for spin and DM noise, and this results in the white noise parameters (EFAC and DMEFAC) being overestimated.

\updated{To compare the DMGP model with the DMX model, we analyzed the same simulated dataset using the piecewise-constant DMX model with 20-day-long DMX ranges (500 DMX parameters in total) instead of the DMGP model.
The DMX parameters are assumed to have infinitely wide Gaussian priors and are analytically marginalized along with the other timing model parameters, following the usual practice in PTA analyses.
To avoid parameter degeneracy, we keep the average DM and its derivative fixed in this analysis, unlike the DMGP analysis.
The prior distributions used in this analysis are listed in Table \ref{tab:sim-params}, and the posterior distribution obtained from this analysis is overplotted in Figure \ref{fig:sim-result}.
We find that the estimates for EFAC and the achromatic red noise spectral parameters are consistent with the ones obtained while using the DMGP model, but the DMEFAC is underestimated when using the DMX model.
This is because the DMX parameters characterize short-timescale DM variability, and therefore can absorb random short-timescale fluctuations caused by the DM measurement noise.
}


\begin{table*}
\begin{tabular}{c|l|c|c}
\hline
\textbf{Parameter} & \textbf{Description \& Unit}                     & \textbf{Simulated value}   & \textbf{Prior distribution} \\\hline \hline
F0                 & Spin frequency (Hz)                              & 100                        &                           AM  \\
F1                 & Spin frequency derivative (Hz/s)                 & $-10^{-15}$                &                         AM    \\
PHOFF                 & Overall phase offset                 & 0                &                             AM \\
RAJ                & Right ascension (hour angle)                     & 05:00:00                   &                             AM \\
DECJ               & Declination (degree)                             & 15:00:00                   &                            AM  \\
DM  $\dag$               & Dispersion measure (pc/cm$^3$)                      & 15                         &                             AM \\
DM1 $\dag$                & Dispersion measure derivative (pc/cm$^3$/yr)        & $10^{-3}$     &                             AM \\
EFAC               & Global EFAC                                      & 1                          & Uniform{[}0.5, 2.0{]}       \\
DMEFAC             & Global DMEFAC                                    & 1                          & Uniform{[}0.5, 2.0{]}       \\
TNREDAMP           & Log-amplitude of the spin noise                  & -13                        & Uniform{[}-16, -11{]}       \\
TNREDGAM           & Spectral index of the spin noise                 & 4                          & Uniform{[}1, 7{]}           \\
TNDMAMP    $\dag$        & Log-amplitude of the DM noise                    & -13                        & Uniform{[}-16, -11{]}       \\
TNDMGAM    $\dag$        & Spectral index of the DM noise                   & 3                          & Uniform{[}1, 7{]}   \\  
DMX\_   $\ddag$         & Piecewise-constant DM values in time                   &                          & AM   \\ \hline
\end{tabular}
\caption{The timing \& noise parameters used to generate the simulated dataset and the prior distributions for noise parameters. 
The timing model parameters are assumed to have Gaussian prior distributions with infinite width and are analytically marginalized \updated{(these are indicated as `AM' in the table).
`$\dag$' indicates parameters which are analytically marginalized or sampled in the DMGP analysis only, and `$\ddag$' indicates parameters which are analytically marginalized in the DMX analysis only.
DM and DM1 are analytically marginalized in the DMGP analysis, but are kept constant in the DMX analysis.
Note that the DMX\_ parameters were not used for the simulation.}}
\label{tab:sim-params}
\end{table*}

\section{Summary}
\label{sec:summary}

We derived a general marginalized likelihood function that can be used to perform single-pulsar noise analysis on wideband datasets with arbitrary DM variation models, including Gaussian process models (Section \ref{sec:linear-timing-model}).
We demonstrated the application of our marginalized likelihood using a simulated dataset (Section \ref{sec:example}), where the likelihood function was implemented with the help of \pint{}.
We showed that our likelihood function is a generalization of the likelihood function given in \citet{AlamArzoumanian+2021} (Appendix \ref{sec:lnlike-marg-dmx}), and derived a factorized version of our likelihood function that can be implemented in \enterprise{} (Appendix \ref{sec:lnlike-enterprise}).
We also derived a general maximum-likelihood estimator for timing model parameters applicable to wideband datasets, which can be incorporated into interactive pulsar timing packages like \pint{} (Appendix \ref{sec:maxlike}).
{Finally,} it is possible to incorporate our results into PTA analysis so that DM variations can be analytically marginalized while searching for cross-pulsar correlations such as those induced by gravitational waves in wideband datasets.

\section*{Acknowledgements}

We thank Boris Goncharov for his comments on the manuscript.

\section*{Software}
\pint{} \citep{LuoRansom+2021,SusobhananKaplan+2024},
\enterprise{} \citep{EllisVallisneri+2020,JohnsonMeyers+2024},
\texttt{emcee} \citep{Foreman-MackeyHogg2013},
\texttt{numpy} \citep{HarrisMillman+2020},
\texttt{scipy} \citep{VirtanenGommers+2020},
\texttt{astropy} \citep{RobitailleTollerud+2013},
\texttt{matplotlib} \citep{Hunter2007},
\texttt{corner} \citep{Foreman-Mackey2016},
\texttt{GetDist} \citep{Lewis2019}.

\section*{Data Availability}

The simulated dataset described in Section \ref{sec:example} and the Jupyter notebook used to analyze it are provided as supplementary materials.



\bibliographystyle{mnras}
\bibliography{dmgp-wb} 




\appendix

\newcommand{\alametal}{\citet{AlamArzoumanian+2021}}

\section{Comparison with the wideband likelihood function of \texorpdfstring{\protect\alametal}{Alam et al. 2021}}
\label{sec:lnlike-marg-dmx}

In Section \ref{sec:linear-timing-model}, we derived the marginalized likelihood for the linearized timing model assuming a general DM variability model.
To derive a special case of this likelihood function assuming a piecewise constant model of the DM (the DMX model), we rewrite equation \eqref{eq:dmn-fourier} as follows.
\begin{equation}
    \mathcal{D}_{\text{DMN}}(t_{i})	=\sum_{j=1}^{N_{\text{dmx}}}
    a_{j}^{\text{DMX}} X_{ij} \,,
\label{eq:dmn-dmx}
\end{equation}
where $X_{ij}$ is 1 if the $i$th TOA and DM measurement falls within the $j$th DMX range, and 0 otherwise.
We assume the DMX ranges are exclusive and exhaustive so that the above basis is orthogonal and covers all TOAs.
Further, when the DMX model is used, other DM model components such as the Taylor series representation of the DM ($D_0$, $D_1$, ...) are not treated as free parameters to avoid parameter degeneracies.

The wideband likelihood given in Appendix B of \citet{AlamArzoumanian+2021} (equations B5 and B8) can be written in our notation as
\begin{align}
\ln\Lambda&=-\frac{1}{2}(\textbf{r}-\boldsymbol{\Delta}_{\text{DMN}})^{T}\textbf{C}_{\text{toa}}^{-1}(\textbf{r}-\boldsymbol{\Delta}_{\text{DMN}})-\frac{1}{2}\ln\det[2\pi\textbf{C}_{\text{toa}}]\nonumber\\
&\qquad-\frac{1}{2}(\boldsymbol{\delta}-\boldsymbol{\mathcal{D}}_{\text{DMN}}-\boldsymbol{\mathcal{D}}'_{\text{jump}})^{T}\textbf{N}_{\text{dm}}^{-1}(\boldsymbol{\delta}-\boldsymbol{\mathcal{D}}_{\text{DMN}}-\boldsymbol{\mathcal{D}}'_{\text{jump}})\nonumber\\
&\qquad-\frac{1}{2}\ln\det[2\pi\textbf{N}_{\text{dm}}]\,,
\label{eq:lnlike-dmx}
\end{align}
where $\boldsymbol{\mathcal{D}}_{\text{DMN}}$ is a vector containing DMs computed using equation \eqref{eq:dmn-dmx},
$\boldsymbol{\Delta}_{\text{DMN}}$ is a vector containing the corresponding dispersion delays computed using equation \eqref{eq:dm-delay},
$\textbf{C}_{\text{toa}}=\textbf{N}_{\text{toa}} + \textbf{M}_{\text{toa}}^T\boldsymbol{\Phi}\textbf{M}_{\text{toa}}$, $\textbf{N}_{\text{toa}}$ and $\textbf{N}_{\text{dm}}$ are the portions of $\textbf{N}$ containing only TOA variances and DM variances respectively, and $\textbf{M}_{\text{toa}}$ is the portion of $\textbf{M}$ containing only derivatives of TOA residuals \textit{excluding derivatives with respect to DMX parameters}. 
Here, the DMX parameters appearing in $\boldsymbol{\mathcal{D}}_{\text{DMN}}$ and the DMJUMPs appearing in $\boldsymbol{\mathcal{D}}'_{\text{jump}}$ are \textit{not} analytically marginalized.

It is straightforward to see that equation \eqref{eq:lnlike-dmx} has a different structure than equation \eqref{eq:lnlike-marg}, namely the former can be separated into a TOA part and a DM part whereas the latter contains TOA-DM cross terms.
This happens purely because DMX and DMJUMP parameters in equation \eqref{eq:lnlike-dmx} are un-marginalized, and marginalizing these parameters will convert equation \eqref{eq:lnlike-dmx} to a form identical to equation \eqref{eq:lnlike-marg}.
Note that \citet{AlamArzoumanian+2021} interprets the DM part of equation \eqref{eq:lnlike-dmx} (their equation B5) as a prior on the DMX parameters, although their expressions are mathematically equivalent to ours.
Our formalism is more general than \citet{AlamArzoumanian+2021} in the sense that it provides a general framework for handling arbitrary models of DM variations and either as deterministic signals or as Gaussian processes, and can incorporate measurements where $\left\langle t_i d_i \right\rangle\ne 0$.

\section{Factorized form of the wideband likelihood function}
\label{sec:lnlike-enterprise}

While equation \eqref{eq:lnlike-marg} is elegant and has the same form as the narrowband likelihood function, it is difficult to implement within \enterprise{} since it was mainly designed to handle narrowband datasets.
In this appendix, we derive an alternative form of the wideband likelihood function which can be easily implemented within \enterprise{}.
This derivation is less general, assuming the measurement covariances $\left\langle t_i d_i \right\rangle=0$. 

We begin by writing down the linearized timing model as
\begin{subequations}
\begin{align}
\textbf{r}&=\textbf{r}'+\textbf{F}\boldsymbol{\alpha}+\textbf{S}\textbf{G}\boldsymbol{\beta}\,,\\
\boldsymbol{\delta}&=\boldsymbol{\delta}'+\textbf{H}\boldsymbol{\gamma}+\textbf{G}\boldsymbol{\beta}\,,
\end{align}
\end{subequations}
where \textbf{F} is the TOA design matrix containing entries only for achromatic parameters, \textbf{G} is the DM design matrix containing entries for astrophysical dispersion parameters, \textbf{H} is the DM design matrix containing entries for instrumental DM jumps, 
\begin{equation}
 \textbf{S}=K\, \text{diag}[\nu_{1}^{2}, \nu_{2}^{2}, ...] \,,
\end{equation}
and $\boldsymbol{\alpha}$, $\boldsymbol{\beta}$, and $\boldsymbol{\gamma}$ are the corresponding parameter deviations.
We also define the diagonal TOA and DM covariance matrices $\textbf{K}$ and $\textbf{L}$, and sets of parameters $\textbf{p}$ and $\textbf{q}$ that appear respectively in $\textbf{K}$ and $\textbf{L}$.
The prior distributions on the parameter deviations are given by
\begin{subequations}
\begin{align}
    P[\boldsymbol{\alpha}|\textbf{A}]&=\mathcal{N}[\boldsymbol{\alpha};0,\textbf{A}]\,,\\P[\boldsymbol{\beta}|\textbf{B}]&=\mathcal{N}[\boldsymbol{\beta};0,\textbf{B}]\,,\\P[\boldsymbol{\gamma}|\boldsymbol{\Gamma}]&=\mathcal{N}[\boldsymbol{\gamma};0,\boldsymbol{\Gamma}]\,,
\end{align}
\end{subequations}
where $\mathcal{N}[\textbf{x};\boldsymbol{\mu},\boldsymbol{\Sigma}]$ is a multivariate normal distribution with mean $\boldsymbol{\mu}$ and covariance $\boldsymbol{\Sigma}$.
These matrices and vectors are related to the objects used in section \ref{sec:linear-timing-model} as 
\begin{subequations}
\begin{align}
    \textbf{M}^{T}&=\left[\begin{array}{cc}
\textbf{F} & 0\\
\textbf{S}\textbf{G} & \textbf{G}\\
0 & \textbf{H}
\end{array}\right]\,,\\\textbf{N}&=\left[\begin{array}{cc}
\textbf{K} & 0\\
0 & \textbf{L}
\end{array}\right]\,,\\\textbf{y}&=\left[\begin{array}{c}
\textbf{r}\\
\boldsymbol{\delta}
\end{array}\right]\,,\\\textbf{y}'&=\left[\begin{array}{c}
\textbf{r}'\\
\boldsymbol{\delta}'
\end{array}\right]\,,\\\textbf{a}&=\left[\begin{array}{c}
\boldsymbol{\alpha}\\
\boldsymbol{\beta}\\
\boldsymbol{\gamma}
\end{array}\right]\,,\\\boldsymbol{\Phi}&=\left[\begin{array}{ccc}
\textbf{A} & 0 & 0\\
0 & \textbf{B} & 0\\
0 & 0 & \boldsymbol{\Gamma}
\end{array}\right]\,,\\
\textbf{b}&=\left[\begin{array}{c}
\textbf{p}\\
\textbf{q}
\end{array}\right]\,.
\end{align}
\end{subequations}

Since the TOA and DM measurements are uncorrelated, we may factorize the likelihood as
\begin{align}
P(\textbf{r},\boldsymbol{\delta}|\boldsymbol{\alpha},\boldsymbol{\beta},\boldsymbol{\gamma},\textbf{p},\textbf{q},\mathfrak{M})&=P(\textbf{r}|\boldsymbol{\alpha},\boldsymbol{\beta},\textbf{p},\mathfrak{M})\;P(\boldsymbol{\delta}|\boldsymbol{\beta},\boldsymbol{\gamma},\textbf{q},\mathfrak{M})\,,
\end{align}
where TOA and DM parts of the likelihood are given by
\begin{subequations}
\begin{align}
P(\textbf{r}|\boldsymbol{\boldsymbol{\alpha}},\boldsymbol{\boldsymbol{\beta}},\textbf{p},\mathfrak{M})&=\mathcal{N}[\textbf{r};(\textbf{F}\boldsymbol{\alpha}+\textbf{S}\textbf{G}\boldsymbol{\beta}),\textbf{K}]\,,\\
P(\boldsymbol{\delta}|\boldsymbol{\boldsymbol{\beta}},\boldsymbol{\gamma},\textbf{q},\mathfrak{M})&=\mathcal{N}[\boldsymbol{\delta};(\textbf{H}\boldsymbol{\gamma}+\textbf{G}\boldsymbol{\beta}),\textbf{L}]\,.
\end{align}
\end{subequations}

Since the parameter deviations $\alpha$ and $\gamma$ are not common between the TOA and DM parts of the likelihood, we can analytically marginalize them separately, and we get
\begin{subequations}
\begin{align}
P(\textbf{r}|\textbf{A},\boldsymbol{\boldsymbol{\beta}},\textbf{p},\mathfrak{M})&=\mathcal{N}[\textbf{r};\textbf{S}\textbf{G}\boldsymbol{\beta},\textbf{K}']\,,\\P(\boldsymbol{\delta}|\boldsymbol{\boldsymbol{\beta}},\boldsymbol{\Gamma},\textbf{q},\mathfrak{M})&=\mathcal{N}[\boldsymbol{\delta};\textbf{G}\boldsymbol{\beta},\textbf{L}']\,,
\end{align}
\end{subequations}
where $\textbf{K}'=\textbf{K}+\textbf{F} \textbf{A} \textbf{F}^T$ and $\textbf{L}'=\textbf{L}+\textbf{G} \boldsymbol{\Gamma} \textbf{G}^T$.

The joint likelihood can be written as
\begin{align}
P(\textbf{r},\boldsymbol{\delta}|\textbf{A},\boldsymbol{\boldsymbol{\beta}},\boldsymbol{\Gamma},\textbf{p},\textbf{q},\mathfrak{M})=\mathcal{N}[\textbf{y};\textbf{M}'\boldsymbol{\beta},\textbf{N}']\,,
\end{align}
where 
\begin{subequations}
\begin{align}
    \textbf{M}'^T&=[\textbf{S}\textbf{G}\quad \textbf{G}]\,,\\  \textbf{N}'&=\begin{bmatrix}\textbf{K}' & 0\\
0 & \textbf{L}'
\end{bmatrix}\,.
\end{align}
\end{subequations}
Marginalizing this over $\boldsymbol{\beta}$, we obtain
\begin{align}
P(\textbf{r},\boldsymbol{\delta}|\textbf{A},\boldsymbol{\textbf{B}},\boldsymbol{\Gamma},\textbf{p},\textbf{q},\mathfrak{M})=\mathcal{N}[\textbf{y};0,\textbf{C}]\,.
\label{eq:lnlike-marg-joint}
\end{align}
Here, $\textbf{C}=\textbf{N}' + \textbf{M}' \textbf{B}\textbf{M}'^T$ is the same matrix as the one given in equation \eqref{eq:cov-marg}.

We can factorize this likelihood as
\begin{align}
P(\textbf{r},\boldsymbol{\delta}|\textbf{A},\boldsymbol{\textbf{B}},\boldsymbol{\Gamma},\textbf{p},\textbf{q},\mathfrak{M})&=P(\boldsymbol{\delta}|\boldsymbol{\textbf{B}},\boldsymbol{\Gamma},\textbf{q},\mathfrak{M})\nonumber\\
&\quad\times P(\textbf{r}|\boldsymbol{\delta},\textbf{A},\boldsymbol{\textbf{B}},\boldsymbol{\Gamma},\textbf{p},\textbf{q},\mathfrak{M})\,.
\end{align}
The first factor is the DM-only marginalized likelihood function
\begin{align}
P(\boldsymbol{\delta}|\boldsymbol{\textbf{B}},\boldsymbol{\Gamma},\textbf{q},\mathfrak{M})=\mathcal{N}[\boldsymbol{\delta};0,\bar{\textbf{L}}]\,,
\end{align}
where $\bar{\textbf{L}}=\textbf{L}'+\textbf{G}\textbf{B}\textbf{G}^T$, and the second factor is the conditional probability of the TOA residuals $\textbf{r}$ given DM residuals $\boldsymbol{\delta}$.
Since the joint distribution given in equation \eqref{eq:lnlike-marg-joint} is Gaussian, this conditional distribution is also Gaussian, and it turns out to be
\begin{align}
P(\textbf{r}|\boldsymbol{\delta},\textbf{A},\boldsymbol{\textbf{B}},\boldsymbol{\Gamma},\textbf{p},\textbf{q},\mathfrak{M})=\mathcal{N}[\textbf{r};\hat{\textbf{r}}_{\delta},\hat{\textbf{C}}_{\delta}]\,,
\end{align}
where
\begin{subequations}
\begin{align}
\hat{\textbf{r}}_{\delta}&=\textbf{S}\textbf{G}\textbf{B}\textbf{G}^T \bar{\textbf{L}}^{-1}\boldsymbol{\delta}\,,\\
\hat{\textbf{C}}_{\delta}&=\textbf{K}'+\textbf{S}\textbf{G}\left(\textbf{B}-\textbf{B}\textbf{G}^{T}\bar{\textbf{L}}^{-1}\textbf{G}\textbf{B}\right)\textbf{G}^{T}\textbf{S}\,.
\end{align}
\end{subequations}
Here, $\hat{\textbf{r}}_{\delta}$ is the dispersion delay corresponding to the maximum-likelihood dispersion parameters estimated using the DM measurements alone.
We have also verified numerically that the above results are equivalent to equation \eqref{eq:lnlike-marg}.

The \texttt{WidebandTimingModel} class in \enterprise{} implements a version of this factorized likelihood specialized for the DMX model with un-marginalized wideband DM jumps.

\section{Linear fitting of wideband data}
\label{sec:maxlike}

Interactive pulsar timing involves the linear maximum-likelihood fitting of timing model parameters while assuming some fixed value of the noise parameters. 
Since previous studies involving wideband datasets had all used the DMX model to account for DM variations, they implemented this linear fitting by maximizing the likelihood function given in equation \eqref{eq:lnlike-dmx}.
In this appendix, we generalize this to incorporate arbitrary variability models.

We begin by rewriting equation \eqref{eq:lnlike-linear} as
\begin{align}
    \ln L&=-\frac{1}{2}\left(\textbf{y}-\textbf{U}\boldsymbol{\alpha}-\textbf{V}\boldsymbol{\beta}\right)^{T}\textbf{N}^{-1}\left(\textbf{y}-\textbf{U}\boldsymbol{\alpha}-\textbf{V}\boldsymbol{\beta}\right)\nonumber\\
    &\quad-\ln\det [2\pi\textbf{N}]\,,
    \label{eq:lnlike-split}
\end{align}
where we have split the parameter deviation vector $\textbf{a}$ into the correlated noise amplitude vector $\boldsymbol{\alpha}$ and the timing model parameter deviation vector $\boldsymbol{\beta}$.
Similarly, the design matrix $\textbf{M}$ is split column-wise into the correlated noise basis matrix $\textbf{U}$ and the timing model design matrix $\textbf{V}$.
We are interested in estimating $\boldsymbol{\beta}$ whereas $\boldsymbol{\alpha}$ are nuisance parameters.
Similar to Section \ref{sec:linear-timing-model}, we impose the prior $\ln\Pi[\boldsymbol{\alpha}]=-\frac{1}{2}\boldsymbol{\alpha}^T\boldsymbol{\Psi}^{-1}\boldsymbol{\alpha}$ where $\boldsymbol{\Psi}$ is a diagonal covariance matrix containing the correlated noise weights (it is the portion of $\boldsymbol{\Phi}$ corresponding to the correlated noise amplitudes).
This prior allows us to marginalize equation \eqref{eq:lnlike-split} over $\boldsymbol{\alpha}$, and we get
\begin{align}
    \ln \Lambda'=-\frac{1}{2}\left(\textbf{y}-\textbf{V}\boldsymbol{\beta}\right)^{T}\textbf{G}^{-1}\left(\textbf{y}-\textbf{V}\boldsymbol{\beta}\right)-\ln\det [2\pi\textbf{G}]\,,
    \label{eq:lnlike-split-marg}
\end{align}
where $\textbf{G}=\textbf{N} + \textbf{U}^T \boldsymbol{\Psi}^{-1} \textbf{U}$.
We now maximize this equation over $\boldsymbol{\beta}$ to obtain its maximum likelihood estimator
\begin{equation}
    \hat{\boldsymbol{\beta}} = \left(\textbf{V}^T\textbf{G}^{-1}\textbf{V}\right)^{-1} \textbf{V}^T \textbf{G}^{-1} \textbf{y}\,.
\end{equation}
The corresponding parameter covariance matrix is 
\begin{align}
    \boldsymbol{K}_\beta = \frac{1}{2N_{\text{toa}}-q} (\bar{\textbf{y}}^T \textbf{G}^{-1} \bar{\textbf{y}}) (\textbf{V}^T \textbf{G}^{-1} \textbf{V})^{-1}\,,
    \label{eq:Kbeta}
\end{align}
where $q$ is the number of timing model parameters and $\bar{\textbf{y}}=\textbf{y}-\textbf{V}\hat{\boldsymbol{\beta}}$ contains the post-fit timing residuals.

Whitened residuals ${\textbf{y}}_w$ can be obtained by maximizing equation \eqref{eq:lnlike-linear} over $\textbf{a}$ and subtracting the resulting signal from the pre-fit residuals, i.e., 
\begin{align}
    \hat{\textbf{a}} &=  \left(\boldsymbol{\Phi}^{-1} + \textbf{M}^T\textbf{N}^{-1}\textbf{M}\right)^{-1} \textbf{M}^T \textbf{N}^{-1} \textbf{y}\,,
    \label{eq:ahat}\\
    {\textbf{y}}_w &= \textbf{y} - \textbf{M} \hat{\textbf{a}}\,
    .
    \label{eq:white-res}
\end{align}


\bsp	
\label{lastpage}
\end{document}